\documentclass[twocolumn,preprintnumbers,superscriptaddress,nofootinbib,aps,prd,floatfix]{revtex4}
\pdfoutput=1

\usepackage{amsmath}
\usepackage{wasysym}
\usepackage{graphicx}
\usepackage{color,array,subfigure,slashed}
\usepackage[scientific-notation=true]{siunitx}
\usepackage[dvipsnames]{xcolor}

\newcommand{\be}{\begin{eqnarray}}
\newcommand{\ee}{\end{eqnarray}}

\newcommand{\bee}{\begin{eqnarray}}
\newcommand{\eee}{\end{eqnarray}}
\newcommand{\beeq}{\begin{equation}}
\newcommand{\eeeq}{\end{equation}}

\numberwithin{equation}{section}

\usepackage{multirow}
\usepackage{array}
\newcolumntype{C}[1]{>{\centering\let\newline\\\arraybackslash\hspace{0pt}}m{#1}}

\begin{document}

\title{Di-Higgs resonance searches in weak boson fusion}

\begin{abstract}
The search for di-Higgs final states is typically limited at the LHC to the dominant gluon fusion channels, with 
weak boson fusion only assuming a spectator role. In this work, we demonstrate that when it comes to
searches for resonant structures that arise from isosinglet mixing in the Higgs sector, the weak boson fusion 
sideline can indeed contribute to winning the discovery game. Extending existing experimental resonance searches 
by including both contributions is therefore crucial.
\end{abstract}

\author{Rahool Kumar Barman} \email{psrkb2284@iacs.res.in}
\affiliation{School of Physical Sciences, Indian Association for the Cultivation of Science, Kolkata 700032, India \\[0.1cm]}
\author{Christoph Englert} \email{christoph.englert@glasgow.ac.uk}
\affiliation{School of Physics and Astronomy, University of
  Glasgow, Glasgow G12 8QQ, United Kingdom\\[0.1cm]}
\author{Dorival Gon\c{c}alves} \email{dorival@okstate.edu}
\affiliation{Department of Physics, Oklahoma State University, Stillwater, Oklahoma, 74078, USA \\[0.1cm]}
\author{Michael Spannowsky} \email{michael.spannowsky@durham.ac.uk}
\affiliation{Institute for Particle Physics Phenomenology, Department
  of Physics, Durham University, Durham DH1 3LE, United Kingdom\\[0.1cm]}

\preprint{IPPP/20/28, OSU-HEP-20-08}

\maketitle

\section{Introduction}
\label{sec:intro}
The search for new physics beyond the Standard Model (SM) is a key pillar of the Large
Hadron Collider (LHC) physics program. As significant deviations from the SM expectation
have remained elusive after the Higgs boson's discovery so far, the nature of the electroweak scale 
is still fundamentally unknown. A particularly relevant process in this context is the production of multiple
Higgs bosons. First, multi-Higgs production directly probes aspects of spontaneous symmetry 
breaking that cannot be accessed with weak boson or heavy quark physics. 
Second, the inclusive production cross section of Higgs pairs of around 30 fb~\cite{Dawson:1998py,Frederix:2014hta,Borowka:2016ehy,deFlorian:2016spz,Grazzini:2018bsd,DiMicco:2019ngk,Baglio:2020ini} is about 3 orders 
of magnitude smaller than single Higgs production, thus highlighting the statistical difficulty that experimental 
investigations face in this area.

Multi-Higgs production is phenomenologically limited to Higgs pairs~\cite{Plehn:2005nk}, at least in the near future~\cite{Papaefstathiou:2015paa,Chiesa:2020awd}, and as with single Higgs production, gluon fusion (GF) contributes to the bulk of the production cross section. While Higgs production via weak boson fusion (WBF) with its distinct phenomenological properties \cite{Cahn:1983ip,Rainwater:1998kj,Rainwater:1999sd,Plehn:1999xi} 
and large cross section plays an important role in the investigation of the Higgs boson's properties, di-Higgs production from 
weak boson fusion will be statistically limited at the LHC~\cite{Dolan:2015zja,Bishara:2016kjn,Arganda:2018ftn}. WBF-type analyses are further hampered by the importance of the top threshold for gluon fusion production~\cite{Dolan:2013rja} and the necessity to relax central jet vetos to retain a reasonable WBF signal count through central $h\to b\bar b$ decays. Experimental analyses typically mitigate the nonapplicability of central jet vetos in the WBF selection by considering stringent invariant jet pair masses; see, e.g., Ref. \cite{Aad:2020kub}. While such a selection serves to purify signal samples toward the WBF component, forward jets will also arise from gluon fusion samples~\cite{DelDuca:2001eu,DelDuca:2003ba,delDuca:2007nwt,DelDuca:2006hk} when biased toward valence quark-flavored initial state processes and the question of the size of the potential, model-dependent GF component remains.

Resonant phenomena in weak boson fusion are less studied from a phenomenological perspective than their GF counterparts. A bias toward GF-like production is understandable as two-Higgs-doublet extensions of the SM in particular as prototypes of supersymmetric theories lead to gauge-phobic scalars, and WBF production of exotic states, e.g., the additional $CP$-odd scalar proceeds dominantly through GF. However, the observation of resonances in WBF would have exciting theoretical implications. Introducing a new resonant beyond the Standard Model scalar in the WBF modes rests on nonalignment~\cite{Grzadkowski:2018ohf}, $CP$ violation~\cite{Fontes:2017zfn}, a significant nondoublet component of the electroweak vacuum~(e.g. \cite{Georgi:1985nv,Gunion:1989ci,Hartling:2014aga}), or combinations of these. 

Electroweak symmetry breaking from triplets faces a theoretical reservation related to the fine-tuning of the rho parameter~\cite{Gunion:1990dt}.\footnote{It remains as a possibility of strong electroweak symmetry breaking in realistic UV constructions~\cite{Ferretti:2014qta,Golterman:2015zwa}.} Phenomenologically, (tree-level) custodial triplet extensions lead to a range of additional exotic final states, most notably a doubly charged Higgs that is predominantly produced through weak boson fusion as part of a fermiophobic custodial quintet~\cite{Godfrey:2010qb,Cheung:2002gd,Englert:2013wga,Zaro:2015ika,Degrande:2015xnm}. Electrically uncharged components of the custodial triplet will not decay promptly to the 125 GeV state if the latter is identified as a doubletlike state, again due to custodial isospin. $CP$ violation is typically a small effect in actual scans~\cite{Fontes:2017zfn} such that a competitive production through WBF is typically suppressed.

The possibility of nonalignment (i.e., the physical 125~GeV Higgs boson not being fully aligned with fluctuations around the electroweak vacuum) remains as an $a$ $priori$ relevant parameter space for WBF to be relevant. The mixing of isospin singlet states is present in any Higgs sector extension, but most transparently analyzed in the so-called Higgs portal scenario~\cite{Binoth:1996au}. This model also fully correlates the exotic Higgs production with observed $m_h\simeq 125$~GeV Higgs boson phenomenology, which turns any sensitivity projection for heavy Higgs states into a conservative estimate as new, nonsinglet fields will loosen the tight correlations of the singlet extensions. 

The relevance of WBF production is further highlighted in singlet scenarios by the fact that for SM Higgs-like states with masses ${\cal{O}}$(TeV), GF and WBF productions become comparable~\cite{Dittmaier:2011ti,Das:2018fog}. This strongly indicates that if such a state is realized in nature, both GF and WBF play $a$ $priori$ an equally important role in the discovery of new physics. 
As there is accidental destructive interference of $pp\to H\to t\bar t$ with QCD continuum top pair production~\cite{Gaemers:1984sj,Dicus:1994bm,Frederix:2007gi,Carena:2016npr,Hespel:2016qaf,BuarqueFranzosi:2017qlm,Englert:2019rga} which particularly affects the sensitivity in the singlet-extension scenario~\cite{Basler:2019nas}, gaining sensitivity in the $H\to hh$ decays is not only necessary but also possibly the only phenomenological robust avenue to successfully detect such scenarios. Depending on the Higgs potential, these channels might be favored over the decays into massive electroweak gauge bosons, which are additional relevant channels.

In this work, we perform a detailed investigation of WBF production of exotic Higgs bosons $pp\to H jj $ arising from isosinglet mixing, in particular in their decay $H\to hh$. We include the gluon fusion component keeping the full $m_t$ dependence and highlight the interplay of both production modes and their relevance to hone the discovery potential at the LHC. In particular, we show that gluon fusion remains phenomenologically relevant and should therefore be reflected as an appropriate signal contribution in any analysis that seeks to inform further theoretical investigations.

We organize this paper as follows: In Sec.~\ref{sec:model}, we provide a short summary of the key phenomenological aspects of the singlet-extension scenario, which acts as the vehicle of this work. We stress that our findings readily generalize to more complex scenarios. Section~\ref{sec:model} is devoted to the WBF di-Higgs resonance analysis. We conclude in Sec.~\ref{sec:conc}.

\section{The model}
\label{sec:model}
We consider the extension of the SM with Higgs doublet $\Phi_s$ by an additional singlet $\Phi_h$ under the SM gauge group
\begin{equation}
V= \mu_s^2 |\Phi_s|^2 + \lambda_s  |\Phi_s|^4 + \mu_h^2 |\Phi_h|^2 + \lambda_h  |\Phi_h|^4 + \eta |\Phi_s|^2   |\Phi_h|^2 \,.
\end{equation}
Expanding around the vacuum expectation values of the respective fields
\begin{equation}
v_i^2={1\over \lambda_i} \left(-\mu_i^2-{\eta \over 2} v_{j\neq i}^2\right),\quad {i,j=s,h}
\end{equation}
via $\Phi_i=(v_i+H_i)/\sqrt{2}$ leads to a mixing of Lagrangian eigenstates in the mass basis
\begin{equation}
\begin{split}
h &= \phantom{-}\cos\theta\, H_s + \sin\theta \,H_h \\
H &= -\sin\theta \, H_s + \cos\theta \, H_h\,.
\end{split}
\end{equation}
We will implicitly identify $h$ with the observed lighter $m_h\simeq 125~\text{GeV}$ boson aligned with the SM expectation; i.e., we will be particularly interested in the region ${\cos\theta\lesssim 1}$. The masses are given by
\begin{equation}
m^2_{h,H} = (\lambda_sv_s^2 +\lambda_h v_h^2) \mp\sqrt{ (\lambda_s v_s^2 - \lambda_h v_h^2)^2 +\eta^2 v_s^2 v_h^2 }\,,
\end{equation}
and 
\begin{equation}
\tan 2\theta = {\eta\,v_s v_h\over \lambda_s v_s^2 - \lambda_h v_h^2}
\end{equation}
while $v_s\simeq 246$~GeV from electroweak symmetry breaking in the SM.

We assume no additional decay channels, which means that signal strengths of the SM-like Higgs are modified $\mu=\cos^2\theta$. 
$H$ boson production cross sections as a function of $m_H$ can be obtained from the SM ones~\cite{Dittmaier:2011ti} by rescaling with $\sin^2\theta$; branching ratios are unmodified for $m_H< 2m_h$. We are particularly interested in the region $m_H\geq 2m_h$ where cascade decays $H\to hh$ are open. In this case, the heavy Higgs partner receives a leading order additional contribution to its decay width
\begin{equation}
\Gamma(H\to hh) = {c_{Hhh}^2 \over 32 m_H \pi}  \sqrt{ 1 - {4 m_h^2 \over m_H^2}} 
\end{equation}
\begin{widetext}
with
\begin{equation}
c_{Hhh} = 3 \sin 2\theta \left(  \lambda_sv_s  \cos \theta - \lambda_hv_h \sin\theta \right)- \tan 2\theta \\
\left( \lambda_s v_s^2- \lambda_h^2 v_h^2 \right) \left[ ( 1-3\cos^2\theta) {\sin\theta \over v_h} - 
( 1-3\sin^2\theta) {\cos\theta \over v_s} \right]\,.
\end{equation}
\end{widetext}
The potential measurement of $\Gamma(H\to hh)$ together with the masses $m_{h,H}$ and SM signal strength and weak boson masses allows us to fully reconstruct the singlet-extended Higgs potential. A range of precision computations from a QCD and electroweak point of view have become available recently \cite{Chen:2014ask,Bojarski:2015kra,Dawson:2017jja,Lopez-Val:2014jva,Falkowski:2015iwa} with strongest constraints typically arising from the $W$ mass measurement~\cite{Lopez-Val:2014jva,Robens:2015gla}.

\section{Analysis}
\label{sec:ana}
We  derive the LHC sensitivity to di-Higgs resonances in the vector boson fusion (VBF) channel $pp \rightarrow Hjj$, with $H\rightarrow h h \rightarrow 4b$. The signal is characterized by four bottom tagged jets  in association with two light-flavor jets. The leading backgrounds for this process  are  $pp\rightarrow 4b + 2j$, $2b + 4j$, and $t\bar{t}b\bar{b}$. 

We generate the WBF and QCD $pp\to (H\to hh) j j$ signal samples with {\sc vbfnlo}~\cite{Arnold:2008rz}, which we have modified to include the $H\to hh$ decay. The backgrounds are generated with {\sc MadGraph5aMC@NLO}~\cite{Alwall:2014hca}. All samples are generated at leading order with center of mass energy of ${\sqrt{s}=13}$~TeV. Parton shower, hadronization, and underlying event effects are accounted for with {\sc pythia8}~\cite{Sjostrand:2007gs}.  Jets are defined through the anti-k$_T$ algorithm with ${R=0.4}$, $p_{Tj}>30$~GeV, and $|\eta_j|<4.5$ via {\sc FastJet}~\cite{Cacciari:2011ma}. We assume $70\%$ $b$-tagging efficiency and a 1\% mistag rate. 

We start our analysis demanding at least six jets in the final state, where four of those are $b$ tagged. We impose a minimum threshold for the invariant mass for the four $b$ jets of $m_{4b}>350$~GeV and veto  leptons with ${p_{T\ell}>12}$~GeV and $|\eta_{\ell}|<2.5$. The two  light-flavor jets with highest rapidity $j_{1,2}$ satisfy the VBF topology falling in different hemispheres of the detector ${\eta_{j1} \times \eta_{j2}<0}$, with large rapidity separation ${|\eta_{j1}- \eta_{j2}|>4.2}$, and  sizable invariant mass ${m_{jj}>1}$~TeV. 

\begin{figure}[t!]
\includegraphics[scale=0.2]{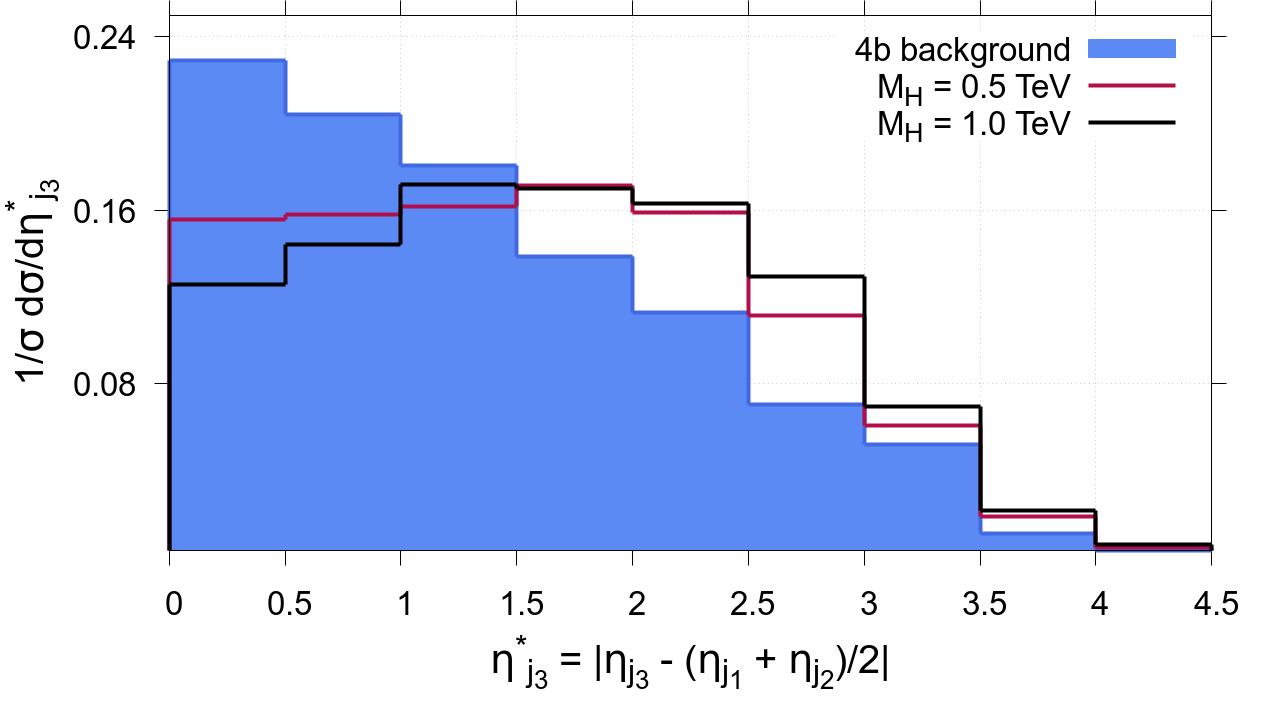}
\caption{Normalized distribution of $\eta_{j_{3}}^{\star}=|\eta_{j3}-(\eta_{j1}+\eta_{j2})/2|$ for the dominant $4b$ background (blue) and the WBF signal events $M_{H} = 0.5~{\rm TeV}$ (red) and  1~TeV (black)  after imposing the basic selection cuts and the VBF selections: ${\eta_{j1} \times \eta_{j2}<0}$, ${|\eta_{j1}- \eta_{j2}|>4.2}$, and ${m_{jj}>1}$~TeV.}
\label{fig:etaj3}
\end{figure}

\begin{table*}[!ht]
\begin{center}
\begin{tabular}{ C{3.1cm} |  C{2.6cm}  C{2.6cm}  C{2.6cm} } 
 Process &  Basic selections & VBF topology & Double Higgs reconstruction \\
\hline 
$4b$ &  250 & 47 & 1.2 \\
$2b2j$ & $4.9 \times 10^{-1}$ & $1.0 \times 10^{-1}$ & - \\
$t\bar{t}b\bar{b}$ & 90 & 3.7 & $3.0 \times 10^{-3}$ \\ \hline
 WBF~${m_{H} = 500~\text{GeV}}$  &  $2.6 \times 10^{-1}$ & $1.3 \times 10^{-1}$ & $5.0 \times 10^{-2}$ \\  
 GF~${m_{H} = 500~\text{GeV}}$  & $2.2 \times 10^{-1}$ & $7.1 \times 10^{-2}$ & $2.8 \times 10^{-2}$ \\ \hline  
 WBF~${m_{H} = 1~\text{TeV}}$  & $9.4 \times 10^{-2}$ & $5.4 \times 10^{-2}$ & $3.2 \times 10^{-2}$ \\  
 GF~${m_{H} = 1~\text{TeV}}$ &  $2.2 \times 10^{-2}$ & $8.3 \times 10^{-3}$ & $4.7\times 10^{-3}$ \\ 
\end{tabular}
\caption{Cut-flow table showing the cross section (in fb) for the VBF signal and backgrounds. The VBF signal is decomposed between the WBF and GF components. The background rates  are normalized by the next-to-leading-order (NLO) K factors: 1.7 $(4b)$~\cite{Alwall:2014hca}, 1.3 $(2b2j)$~\cite{Alwall:2014hca}, and 1.8 $(t\bar{t}b\bar{b})$~\cite{Buccioni:2019plc}. The signal rate is given with  $\text{BR}(H\rightarrow hh)=1$ and $\sin\theta=0.3$. The GF signal rates are also normalized by the NLO K factor: 1.65. QCD corrections for the WBF process are included through an appropriate scale choice~\cite{Figy:2003nv} and
through MCFM for the gluon fusion contribution employing the heavy top limit~\cite{Campbell:2006xx,Campbell:2019dru}.
}
\label{tab:cutflow}
\end{center}
\end{table*}

While the WBF signal displays suppressed extra jet emissions in the central region of the detector, the bulk of the  QCD background radiation is centered around this regime~\cite{Derrick:1987uy,Bjorken:1992er,Barger:1991ar,Barger:1994zq}.  In Fig.~\ref{fig:etaj3}, we illustrate this  property displaying two mass scenarios for the WBF signal samples, $m_H=0.5$ and 1~TeV. The more massive the signal resonance is, the further forward the tagging jets hit the detector. This phenomenological pattern is related to gauge boson scattering $VV\rightarrow hh$ around the heavy Higgs pole, where the  longitudinal and transverse scattering amplitudes scale as $\mathcal{A}_{LL}/\mathcal{A}_{TT}\sim m_{H}^2/m_V^2$ for   $m_H\gg m_V$~\cite{Dawson:1984gx,Figy:2007kv,Goncalves:2017gzy}.   We explore this feature to further suppress the backgrounds imposing that the rapidity for the third jet $\eta_{j3}$ satisfies the relation
\begin{equation}
\left| \eta_{j3}-\frac{\eta_{j1}+\eta_{j2}}{2}\right|>2.5\,.
\end{equation}

After establishing the VBF topology, the next step of the analysis focuses on the Higgs boson's reconstruction. This is performed by identifying among the four $b$ jets the pair whose invariant mass $m_{h1}$ is closest to the Higgs mass, $m_h=125$~GeV. The remaining  $b$-jet pair defines the second Higgs boson candidate $h_2$. In the two-dimensional space defined by the masses of the Higgs boson candidates $(m_{h1},m_{h2})$, the signal region is defined to be within the circular region
\begin{equation}
\sqrt{\left(\frac{m_{h1}-125~\text{GeV}}{20~\text{GeV}}\right)^2+\left(\frac{m_{h2}-125~\text{GeV}}{20~\text{GeV}}\right)^2}<1\,.
\end{equation}

To further improve the $m_{4b}$ mass resolution, each Higgs boson candidate's four-momentum is scaled by the correction factor $m_h/m_{h1(2)}$. This improves the signal $m_{4b}$ resolution from $20\%$ to $40\%$, depending on the heavy Higgs mass  hypothesis, and presents subleading effects to the background $m_{4b}$ distribution~\cite{Sirunyan:2018zkk}.

\begin{figure}[b!]
\includegraphics[scale=0.42]{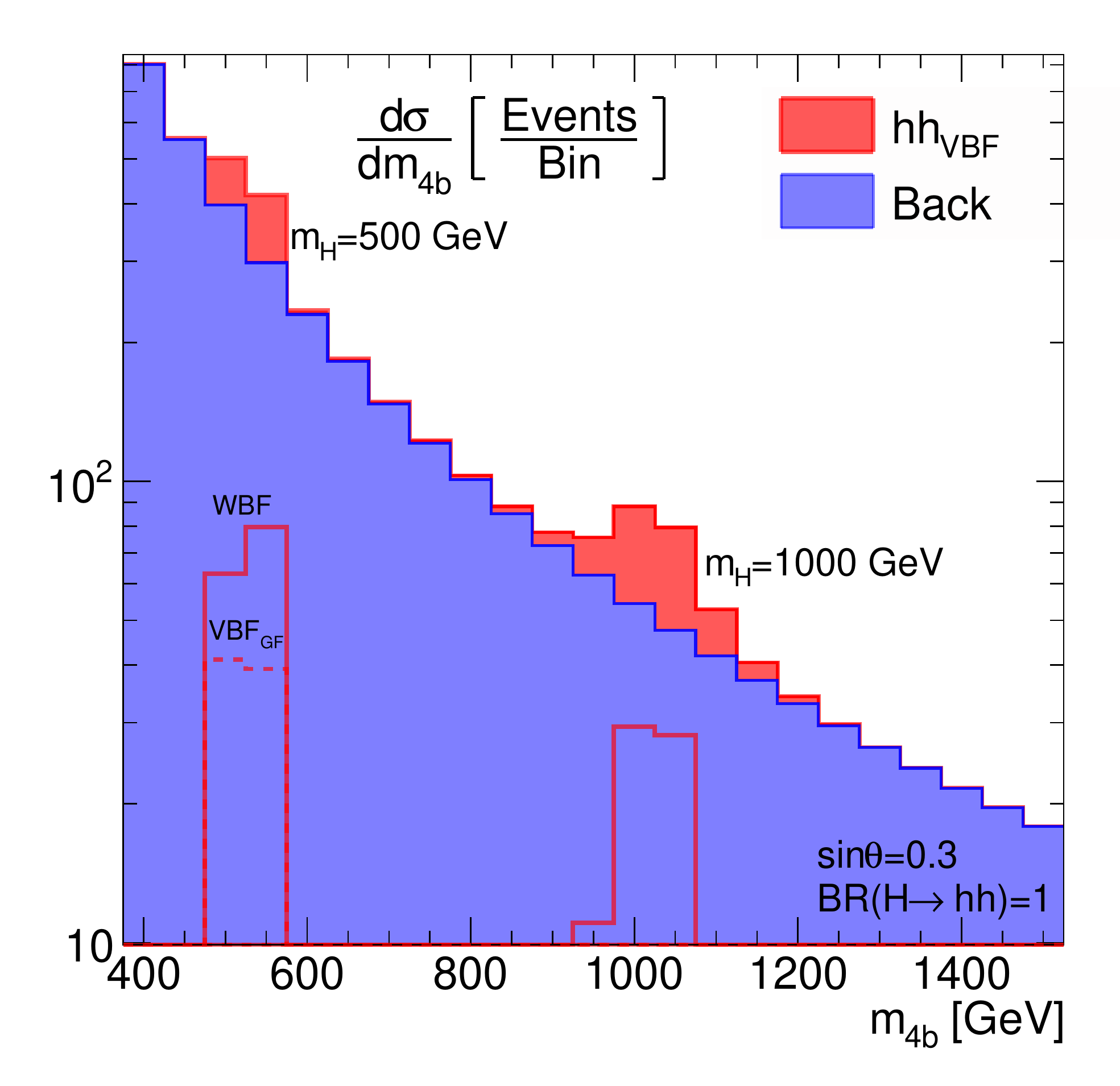}
\caption{Stacked $m_{4b}$ distribution for the  signal and background events after the complete cut-flow analysis shown  in Table~\ref{tab:cutflow}.  The VBF signal hypotheses are also shown in the nonstacked format with the WBF (solid line) and GF (dashed line) components independently displayed. We assume  $\text{BR}(H\rightarrow hh)=1$ and $\sin\theta=0.3$ with the LHC running at  $\sqrt{s}=13~\text{TeV}$ and integrated luminosity $\mathcal{L}=3~\text{ab}^{-1}$.}
\label{fig:mhh}
\end{figure}

Since very few multijet background events pass the cut-flow analysis with large $m_{4b}$, we follow a similar statistical procedure performed by the ATLAS Collaboration in their $pp\rightarrow H\rightarrow hh\rightarrow 4b$ study~\cite{Aaboud:2018knk}. Namely, the statistical precision for the $m_{4b}$ distribution at high energies is improved by fitting the background distribution at low invariant masses  $m_{4b}<1$~TeV with the functional form
\begin{equation}
F(m_{4b})=a\frac{s}{m_{4b}^2}\left(1-\frac{m_{4b}}{\sqrt{s}}\right)^{b-c \log\frac{m_{4b}}{\sqrt{s} }} \,,
\end{equation}
where $a,~b,$ and $c$ are real free parameters and $\sqrt{s}$ the LHC center-of-mass energy. This also 
emulates a data-driven approach that is typically the method of choice when backgrounds are only poorly understood from a systematic and theoretical perspective; see, e.g.,~\cite{Aaboud:2018urx,Aad:2020kop}. As we are looking for a resonance on top of a steeply falling background, such a method provides a particularly motivated approach to reduce uncertainties.

In Fig.~\ref{fig:mhh}, we illustrate the invariant mass distribution $m_{4b}$ for the signal and background components after the full cut-flow analysis shown  in Table~\ref{tab:cutflow}. While the WBF signal component displays dominant contributions to the event rate, the VBF GF signal can result in non-negligible additions to the event count. It should be noted that the larger the signal mass $m_H$ is, the larger the relative WBF component becomes.

\begin{figure}[b!]
\includegraphics[width=8cm]{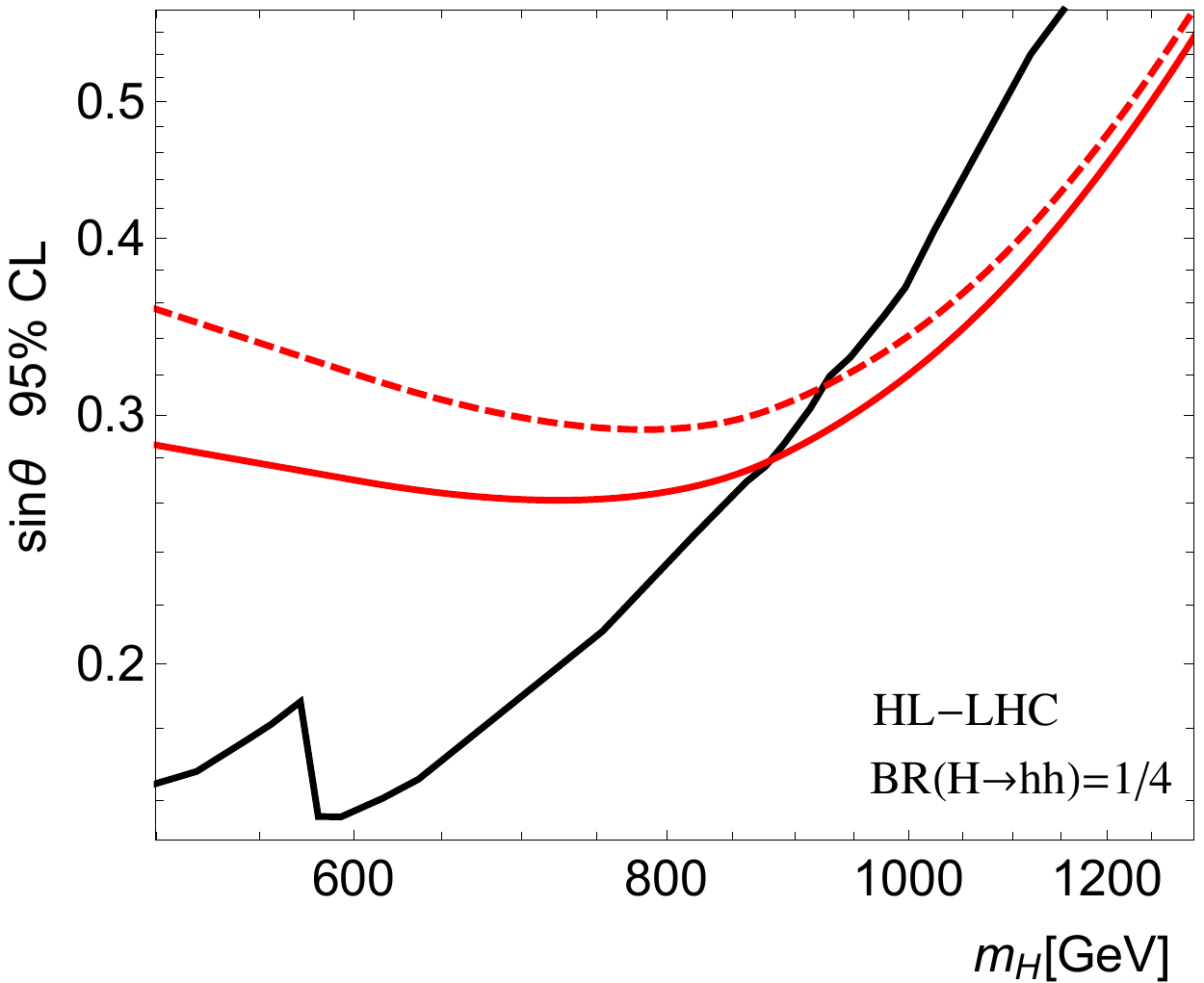}
\caption{95\% C.L. limit on the Higgs-singlet mixing as a function of the heavy Higgs boson mass $m_H$. We show both the  VBF $pp\rightarrow H jj \rightarrow 4bjj$~(red solid) and GF $pp\rightarrow H\rightarrow 4b$ (black) limits. To estimate the importance of the VBF GF signal component to the VBF analysis, we also show  the bound considering only the WBF signal component (red dashed). We assume the heavy Higgs boson branching ratio to di-Higgs  $\text{BR}(H\rightarrow hh) =1/4$ and the LHC at 13~TeV   with integrated luminosity ${\mathcal{L}=3~\text{ab}^{-1}}$.}
\label{fig:limit}
\end{figure}

To estimate the High Luminosity~(HL)-LHC sensitivity to the resonant VBF $hh$ signal, we calculate a binned log-likelihood analysis based on the $m_{4b}$ distribution using the CL$_s$ method~\cite{Read:2002hq}. We assume the integrated luminosity ${\mathcal{L}=3~\text{ab}^{-1}}$. In Fig.~\ref{fig:limit}, we present the 95\% C.L. sensitivity to the heavy Higgs-singlet mixing $\sin\theta$ as a function of the heavy Higgs boson mass $m_H$. Motivated by the Goldstone boson equivalence theorem for $m_H\gg m_W$, we assume the heavy Higgs branching ratio to di-Higgs ${\text{BR}}(H\rightarrow hh) =1/4$. To illustrate the importance of the VBF GF signal component, we  separately show the signal sensitivity accounting for  the full VBF sample  and only for its WBF component. We observe that the VBF GF results in non-negligible contributions for the low mass regime $500~\text{GeV}<m_H<900~\text{GeV}$.

\begin{figure*}[t!]
\subfigure[~]{\includegraphics[width=8cm]{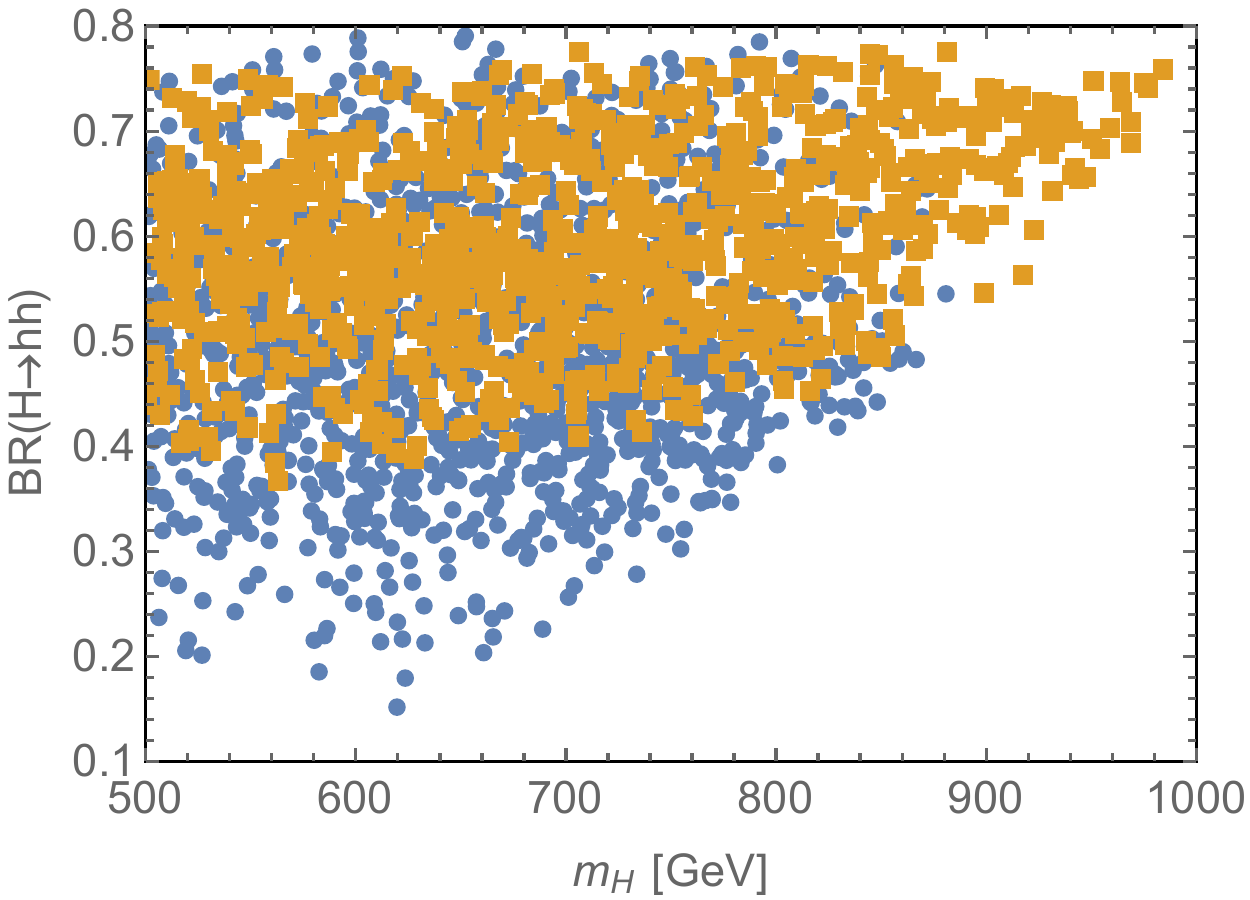}}
\hskip 0.6cm
\subfigure[~]{\includegraphics[width=8cm]{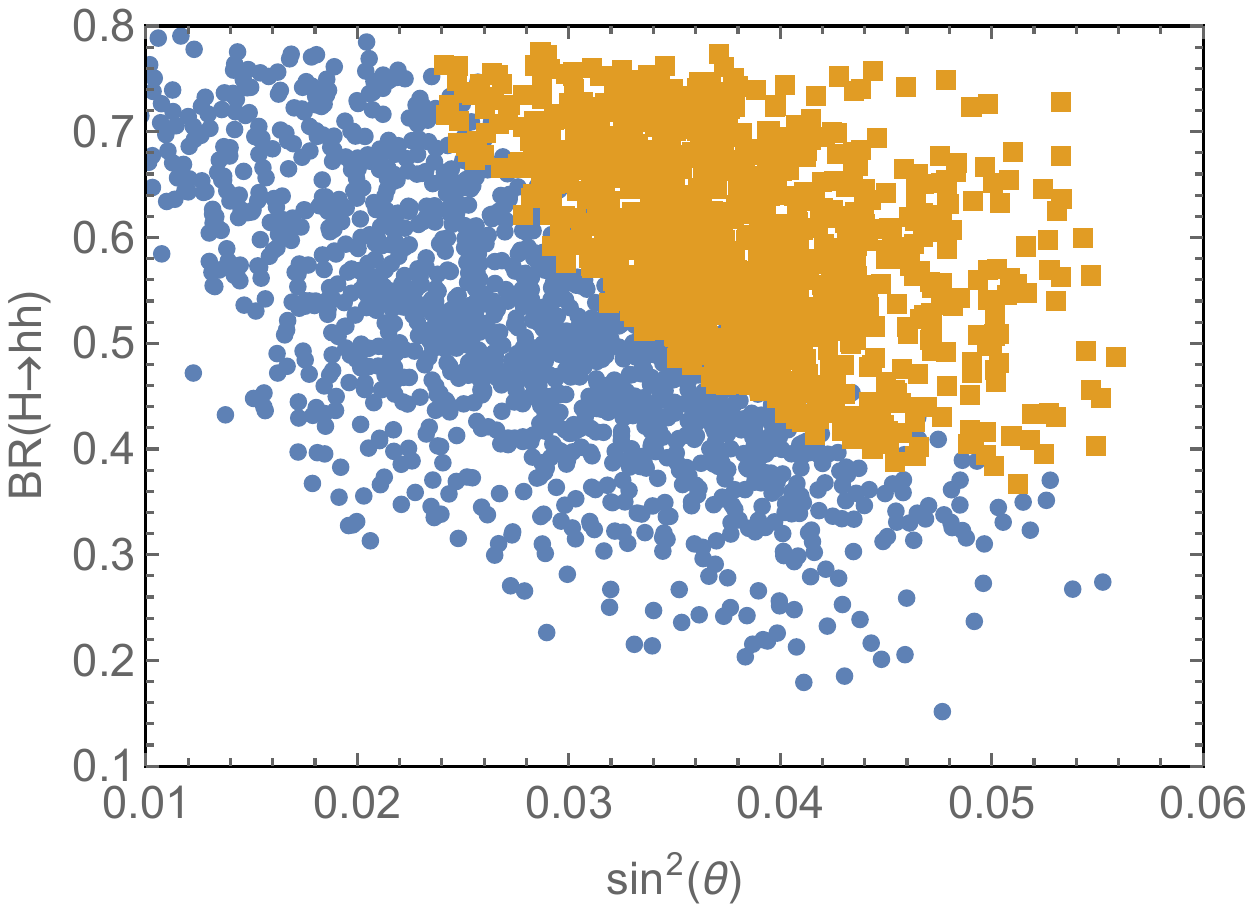}}
\caption{\label{fig:scan} 95\% confidence level constraints interpreted in the singlet scenario of Sec.~\ref{sec:model}. (a) Constraints from gluon fusion in blue dots and from $hh+2j$ in orange squares as a function of mass and branching fraction of $H\to hh$. (b) 
Similar as (a), but we show the correlation of $\sin^2\theta$ with the $H\to hh$ branching.}
\end{figure*}

To compare our new VBF  di-Higgs resonance  search  with the existing  limits, we use the CMS $pp\rightarrow H\rightarrow hh \rightarrow 4b$ study~\cite{Sirunyan:2018zkk}. CMS derives the 95\%~C.L. limit on the heavy Higgs cross section $\sigma(pp\rightarrow H \rightarrow hh \rightarrow 4b)$ as a function of its mass $m_H$. We translate this bound in terms of the mixing $\sin\theta$ in Fig.~\ref{fig:limit}, using the heavy Higgs production cross section at next-to-next-to-leading-order~(NNLO)+ next-to-next-to-leading logarithmic~(NNLL) QCD, including top and bottom quark mass effects up to NLO~\cite{Dittmaier:2011ti,Heinemeyer:2013tqa,deFlorian:2016spz}. The CMS limit on the heavy Higgs cross section was scaled to the HL-LHC integrated luminosity, $\mathcal{L}=3~\text{ab}^{-1}$.  The discontinuity on the CMS limit at $m_H\sim 580$~GeV arises from the two distinct strategies separating  low and high mass resonances.

We observe that the double Higgs resonant search in the VBF mode can significantly contribute to the heavy Higgs resonant analyses. The increase in the ratio $\sigma_\text{VBF}/\sigma_\text{GF}$ for larger $m_H$ leads to comparable sensitivities between the VBF and GF channels for $m_H\sim 900$~GeV. Whereas the VBF search displays stronger limits at the high $m_H$ regime, it can also contribute to further constrain the low mass scenarios ${500~\text{GeV}<m_H<900~\text{GeV}}$ via a combination between  the GF and VBF analyses.

In order to understand the relevance of the GF and VBF limits on the singlet-extension scenario discussed in Sec.~\ref{sec:model}, we interpret the constraints in the aforesaid model. We scan over the singlet model parameter space for $|\lambda_i|\leq 4\pi$ and include the $W$ mass constraint from Refs.~\cite{Lopez-Val:2014jva,Robens:2015gla} as it typically imposes the strongest constraint on the model's parameter space. The results are shown in Fig.~\ref{fig:scan}. The constraints from gluon fusion $gg\to hh$ are displayed in blue points, while those of $pp\to hh j j$ are given in orange squares. We see that the vector boson fusion provides significant sensitivity for higher masses where the gluon fusion projection becomes insensitive. 

While there is a region where gluon fusion and VBF overlap and can be used to further hone the LHC sensitivity to this scenario through a statistical combination, we also see regions in the branching ratio $H\to hh$ where VBF provides genuine, new sensitivity that cannot be accessed with the gluon fusion analysis. This region is characterized by 125 GeV Higgs boson signal strength modifiers of $\lesssim 4\%$. Given the HL-LHC projections of Ref.~\cite{deBlas:2019rxi}, this suggests that the resonance search in the WBF channel can also explore the model's parameter space beyond the precision that can be obtained from 125 GeV signal studies. 

QCD contributions to $pp\to hh j j$ are not the dominant contribution in this mass region (it is a sizable contribution for the theoretical interpretation of the results of Ref.~\cite{Aad:2020kub}), it nonetheless is sizable and should be included in investigations possibly as a separate signal contribution to enable a consistent theoretical interpretation.

\section{Summary and Conclusions}
\label{sec:conc}
Weak boson fusion through its distinct phenomenological properties provides a unique opportunity for new physics
searches. In scenarios with isospin singlet mixing, decays of a heavy Higgs partner into 125 GeV Higgs bosons can be preferred while more obvious decays into top quarks suffer from interference distortion~\cite{Basler:2019nas}, and decays into massive weak bosons might be less dominant. Given that the weak boson fusion production cross section becomes comparable to a gluon fusion cross section for SM-like production at around 1 TeV, the WBF production at small mixing angles becomes a phenomenologically relevant channel. In this paper
we investigate the WBF production of heavy Higgs partners with subsequent decay $H\to h h$. We show that this channel, which has been somewhat overlooked in the past, provides additional relevant new physics potential. In parallel, we show that the gluon fusion component of the vector boson fusion channel remains sizeable and should be included in experimental analysis to enable a consistent theoretical interpretation of reported results.

\bigskip
\noindent{\it{Acknowledgements}} --- 
We thank Stephen Brown and Peter Galler for helpful conversations.

C.E. is supported by the UK Science and Technology Facilities Council (STFC), under Grant No. ST/P000746/1. C.E. also acknowledges support through the IPPP associate scheme. 
D.G. was supported by the U.S.  Department  of  Energy under Grant No. {DE-SC 0016013}. 
M.S. is supported by the STFC under Grant No. ST/P001246/1. 



\begin{thebibliography}{75}
\expandafter\ifx\csname natexlab\endcsname\relax\def\natexlab#1{#1}\fi
\expandafter\ifx\csname bibnamefont\endcsname\relax
  \def\bibnamefont#1{#1}\fi
\expandafter\ifx\csname bibfnamefont\endcsname\relax
  \def\bibfnamefont#1{#1}\fi
\expandafter\ifx\csname citenamefont\endcsname\relax
  \def\citenamefont#1{#1}\fi
\expandafter\ifx\csname url\endcsname\relax
  \def\url#1{\texttt{#1}}\fi
\expandafter\ifx\csname urlprefix\endcsname\relax\def\urlprefix{URL }\fi
\providecommand{\bibinfo}[2]{#2}
\providecommand{\eprint}[2][]{\url{#2}}

\bibitem[{\citenamefont{Dawson et~al.}(1998)\citenamefont{Dawson, Dittmaier,
  and Spira}}]{Dawson:1998py}
\bibinfo{author}{\bibfnamefont{S.}~\bibnamefont{Dawson}},
  \bibinfo{author}{\bibfnamefont{S.}~\bibnamefont{Dittmaier}},
  \bibnamefont{and} \bibinfo{author}{\bibfnamefont{M.}~\bibnamefont{Spira}},
  \bibinfo{journal}{Phys. Rev.} \textbf{\bibinfo{volume}{D58}},
  \bibinfo{pages}{115012} (\bibinfo{year}{1998}), \eprint{hep-ph/9805244}.

\bibitem[{\citenamefont{Frederix et~al.}(2014)\citenamefont{Frederix, Frixione,
  Hirschi, Maltoni, Mattelaer, Torrielli, Vryonidou, and
  Zaro}}]{Frederix:2014hta}
\bibinfo{author}{\bibfnamefont{R.}~\bibnamefont{Frederix}},
  \bibinfo{author}{\bibfnamefont{S.}~\bibnamefont{Frixione}},
  \bibinfo{author}{\bibfnamefont{V.}~\bibnamefont{Hirschi}},
  \bibinfo{author}{\bibfnamefont{F.}~\bibnamefont{Maltoni}},
  \bibinfo{author}{\bibfnamefont{O.}~\bibnamefont{Mattelaer}},
  \bibinfo{author}{\bibfnamefont{P.}~\bibnamefont{Torrielli}},
  \bibinfo{author}{\bibfnamefont{E.}~\bibnamefont{Vryonidou}},
  \bibnamefont{and} \bibinfo{author}{\bibfnamefont{M.}~\bibnamefont{Zaro}},
  \bibinfo{journal}{Phys. Lett. B} \textbf{\bibinfo{volume}{732}},
  \bibinfo{pages}{142} (\bibinfo{year}{2014}), \eprint{1401.7340}.

\bibitem[{\citenamefont{Borowka et~al.}(2016)\citenamefont{Borowka, Greiner,
  Heinrich, Jones, Kerner, Schlenk, Schubert, and Zirke}}]{Borowka:2016ehy}
\bibinfo{author}{\bibfnamefont{S.}~\bibnamefont{Borowka}},
  \bibinfo{author}{\bibfnamefont{N.}~\bibnamefont{Greiner}},
  \bibinfo{author}{\bibfnamefont{G.}~\bibnamefont{Heinrich}},
  \bibinfo{author}{\bibfnamefont{S.~P.} \bibnamefont{Jones}},
  \bibinfo{author}{\bibfnamefont{M.}~\bibnamefont{Kerner}},
  \bibinfo{author}{\bibfnamefont{J.}~\bibnamefont{Schlenk}},
  \bibinfo{author}{\bibfnamefont{U.}~\bibnamefont{Schubert}}, \bibnamefont{and}
  \bibinfo{author}{\bibfnamefont{T.}~\bibnamefont{Zirke}},
  \bibinfo{journal}{Phys. Rev. Lett.} \textbf{\bibinfo{volume}{117}},
  \bibinfo{pages}{012001} (\bibinfo{year}{2016}), \bibinfo{note}{[Erratum:
  Phys. Rev. Lett.117,no.7,079901(2016)]}, \eprint{1604.06447}.

\bibitem[{\citenamefont{de~Florian et~al.}(2016)}]{deFlorian:2016spz}
\bibinfo{author}{\bibfnamefont{D.}~\bibnamefont{de~Florian}}
  \bibnamefont{et~al.} (\bibinfo{collaboration}{LHC Higgs Cross Section Working
  Group}) (\bibinfo{year}{2016}), \eprint{1610.07922}.

\bibitem[{\citenamefont{Grazzini et~al.}(2018)\citenamefont{Grazzini, Heinrich,
  Jones, Kallweit, Kerner, Lindert, and Mazzitelli}}]{Grazzini:2018bsd}
\bibinfo{author}{\bibfnamefont{M.}~\bibnamefont{Grazzini}},
  \bibinfo{author}{\bibfnamefont{G.}~\bibnamefont{Heinrich}},
  \bibinfo{author}{\bibfnamefont{S.}~\bibnamefont{Jones}},
  \bibinfo{author}{\bibfnamefont{S.}~\bibnamefont{Kallweit}},
  \bibinfo{author}{\bibfnamefont{M.}~\bibnamefont{Kerner}},
  \bibinfo{author}{\bibfnamefont{J.~M.} \bibnamefont{Lindert}},
  \bibnamefont{and}
  \bibinfo{author}{\bibfnamefont{J.}~\bibnamefont{Mazzitelli}},
  \bibinfo{journal}{JHEP} \textbf{\bibinfo{volume}{05}}, \bibinfo{pages}{059}
  (\bibinfo{year}{2018}), \eprint{1803.02463}.

\bibitem[{\citenamefont{Alison et~al.}(2019)}]{DiMicco:2019ngk}
\bibinfo{author}{\bibfnamefont{J.}~\bibnamefont{Alison}} \bibnamefont{et~al.},
  in \emph{\bibinfo{booktitle}{{Double Higgs Production at Colliders Batavia,
  IL, USA, September 4, 2018-9, 2019}}}, edited by
  \bibinfo{editor}{\bibfnamefont{B.}~\bibnamefont{Di~Micco}},
  \bibinfo{editor}{\bibfnamefont{M.}~\bibnamefont{Gouzevitch}},
  \bibinfo{editor}{\bibfnamefont{J.}~\bibnamefont{Mazzitelli}},
  \bibnamefont{and} \bibinfo{editor}{\bibfnamefont{C.}~\bibnamefont{Vernieri}}
  (\bibinfo{year}{2019}), \eprint{1910.00012},
  \urlprefix\url{https://lss.fnal.gov/archive/2019/conf/fermilab-conf-19-468-e-t.pdf}.

\bibitem[{\citenamefont{Baglio et~al.}(2020)\citenamefont{Baglio, Campanario,
  Glaus, Mühlleitner, Ronca, Spira, and Streicher}}]{Baglio:2020ini}
\bibinfo{author}{\bibfnamefont{J.}~\bibnamefont{Baglio}},
  \bibinfo{author}{\bibfnamefont{F.}~\bibnamefont{Campanario}},
  \bibinfo{author}{\bibfnamefont{S.}~\bibnamefont{Glaus}},
  \bibinfo{author}{\bibfnamefont{M.}~\bibnamefont{Mühlleitner}},
  \bibinfo{author}{\bibfnamefont{J.}~\bibnamefont{Ronca}},
  \bibinfo{author}{\bibfnamefont{M.}~\bibnamefont{Spira}}, \bibnamefont{and}
  \bibinfo{author}{\bibfnamefont{J.}~\bibnamefont{Streicher}},
  \bibinfo{journal}{JHEP} \textbf{\bibinfo{volume}{04}}, \bibinfo{pages}{181}
  (\bibinfo{year}{2020}), \eprint{2003.03227}.

\bibitem[{\citenamefont{Plehn and Rauch}(2005)}]{Plehn:2005nk}
\bibinfo{author}{\bibfnamefont{T.}~\bibnamefont{Plehn}} \bibnamefont{and}
  \bibinfo{author}{\bibfnamefont{M.}~\bibnamefont{Rauch}},
  \bibinfo{journal}{Phys. Rev.} \textbf{\bibinfo{volume}{D72}},
  \bibinfo{pages}{053008} (\bibinfo{year}{2005}), \eprint{hep-ph/0507321}.

\bibitem[{\citenamefont{Papaefstathiou and
  Sakurai}(2016)}]{Papaefstathiou:2015paa}
\bibinfo{author}{\bibfnamefont{A.}~\bibnamefont{Papaefstathiou}}
  \bibnamefont{and} \bibinfo{author}{\bibfnamefont{K.}~\bibnamefont{Sakurai}},
  \bibinfo{journal}{JHEP} \textbf{\bibinfo{volume}{02}}, \bibinfo{pages}{006}
  (\bibinfo{year}{2016}), \eprint{1508.06524}.

\bibitem[{\citenamefont{Chiesa et~al.}(2020)\citenamefont{Chiesa, Maltoni,
  Mantani, Mele, Piccinini, and Zhao}}]{Chiesa:2020awd}
\bibinfo{author}{\bibfnamefont{M.}~\bibnamefont{Chiesa}},
  \bibinfo{author}{\bibfnamefont{F.}~\bibnamefont{Maltoni}},
  \bibinfo{author}{\bibfnamefont{L.}~\bibnamefont{Mantani}},
  \bibinfo{author}{\bibfnamefont{B.}~\bibnamefont{Mele}},
  \bibinfo{author}{\bibfnamefont{F.}~\bibnamefont{Piccinini}},
  \bibnamefont{and} \bibinfo{author}{\bibfnamefont{X.}~\bibnamefont{Zhao}}
  (\bibinfo{year}{2020}), \eprint{2003.13628}.

\bibitem[{\citenamefont{Cahn and Dawson}(1984)}]{Cahn:1983ip}
\bibinfo{author}{\bibfnamefont{R.~N.} \bibnamefont{Cahn}} \bibnamefont{and}
  \bibinfo{author}{\bibfnamefont{S.}~\bibnamefont{Dawson}},
  \bibinfo{journal}{Phys. Lett.} \textbf{\bibinfo{volume}{136B}},
  \bibinfo{pages}{196} (\bibinfo{year}{1984}), \bibinfo{note}{[Erratum: Phys.
  Lett.138B,464(1984)]}.

\bibitem[{\citenamefont{Rainwater et~al.}(1998)\citenamefont{Rainwater,
  Zeppenfeld, and Hagiwara}}]{Rainwater:1998kj}
\bibinfo{author}{\bibfnamefont{D.~L.} \bibnamefont{Rainwater}},
  \bibinfo{author}{\bibfnamefont{D.}~\bibnamefont{Zeppenfeld}},
  \bibnamefont{and} \bibinfo{author}{\bibfnamefont{K.}~\bibnamefont{Hagiwara}},
  \bibinfo{journal}{Phys. Rev.} \textbf{\bibinfo{volume}{D59}},
  \bibinfo{pages}{014037} (\bibinfo{year}{1998}), \eprint{hep-ph/9808468}.

\bibitem[{\citenamefont{Rainwater and Zeppenfeld}(1999)}]{Rainwater:1999sd}
\bibinfo{author}{\bibfnamefont{D.~L.} \bibnamefont{Rainwater}}
  \bibnamefont{and}
  \bibinfo{author}{\bibfnamefont{D.}~\bibnamefont{Zeppenfeld}},
  \bibinfo{journal}{Phys. Rev.} \textbf{\bibinfo{volume}{D60}},
  \bibinfo{pages}{113004} (\bibinfo{year}{1999}), \bibinfo{note}{[Erratum:
  Phys. Rev.D61,099901(2000)]}, \eprint{hep-ph/9906218}.

\bibitem[{\citenamefont{Plehn et~al.}(2000)\citenamefont{Plehn, Rainwater, and
  Zeppenfeld}}]{Plehn:1999xi}
\bibinfo{author}{\bibfnamefont{T.}~\bibnamefont{Plehn}},
  \bibinfo{author}{\bibfnamefont{D.~L.} \bibnamefont{Rainwater}},
  \bibnamefont{and}
  \bibinfo{author}{\bibfnamefont{D.}~\bibnamefont{Zeppenfeld}},
  \bibinfo{journal}{Phys. Rev.} \textbf{\bibinfo{volume}{D61}},
  \bibinfo{pages}{093005} (\bibinfo{year}{2000}), \eprint{hep-ph/9911385}.

\bibitem[{\citenamefont{Dolan et~al.}(2015)\citenamefont{Dolan, Englert,
  Greiner, Nordstrom, and Spannowsky}}]{Dolan:2015zja}
\bibinfo{author}{\bibfnamefont{M.~J.} \bibnamefont{Dolan}},
  \bibinfo{author}{\bibfnamefont{C.}~\bibnamefont{Englert}},
  \bibinfo{author}{\bibfnamefont{N.}~\bibnamefont{Greiner}},
  \bibinfo{author}{\bibfnamefont{K.}~\bibnamefont{Nordstrom}},
  \bibnamefont{and}
  \bibinfo{author}{\bibfnamefont{M.}~\bibnamefont{Spannowsky}},
  \bibinfo{journal}{Eur. Phys. J.} \textbf{\bibinfo{volume}{C75}},
  \bibinfo{pages}{387} (\bibinfo{year}{2015}), \eprint{1506.08008}.

\bibitem[{\citenamefont{Bishara et~al.}(2017)\citenamefont{Bishara, Contino,
  and Rojo}}]{Bishara:2016kjn}
\bibinfo{author}{\bibfnamefont{F.}~\bibnamefont{Bishara}},
  \bibinfo{author}{\bibfnamefont{R.}~\bibnamefont{Contino}}, \bibnamefont{and}
  \bibinfo{author}{\bibfnamefont{J.}~\bibnamefont{Rojo}},
  \bibinfo{journal}{Eur. Phys. J.} \textbf{\bibinfo{volume}{C77}},
  \bibinfo{pages}{481} (\bibinfo{year}{2017}), \eprint{1611.03860}.

\bibitem[{\citenamefont{Arganda et~al.}(2019)\citenamefont{Arganda,
  Garcia-Garcia, and Herrero}}]{Arganda:2018ftn}
\bibinfo{author}{\bibfnamefont{E.}~\bibnamefont{Arganda}},
  \bibinfo{author}{\bibfnamefont{C.}~\bibnamefont{Garcia-Garcia}},
  \bibnamefont{and} \bibinfo{author}{\bibfnamefont{M.~J.}
  \bibnamefont{Herrero}}, \bibinfo{journal}{Nucl. Phys.}
  \textbf{\bibinfo{volume}{B945}}, \bibinfo{pages}{114687}
  (\bibinfo{year}{2019}), \bibinfo{note}{[Nucl. Phys.B945,114687(2019)]},
  \eprint{1807.09736}.

\bibitem[{\citenamefont{Dolan et~al.}(2014)\citenamefont{Dolan, Englert,
  Greiner, and Spannowsky}}]{Dolan:2013rja}
\bibinfo{author}{\bibfnamefont{M.~J.} \bibnamefont{Dolan}},
  \bibinfo{author}{\bibfnamefont{C.}~\bibnamefont{Englert}},
  \bibinfo{author}{\bibfnamefont{N.}~\bibnamefont{Greiner}}, \bibnamefont{and}
  \bibinfo{author}{\bibfnamefont{M.}~\bibnamefont{Spannowsky}},
  \bibinfo{journal}{Phys. Rev. Lett.} \textbf{\bibinfo{volume}{112}},
  \bibinfo{pages}{101802} (\bibinfo{year}{2014}), \eprint{1310.1084}.

\bibitem[{\citenamefont{Aad et~al.}(2020{\natexlab{a}})}]{Aad:2020kub}
\bibinfo{author}{\bibfnamefont{G.}~\bibnamefont{Aad}} \bibnamefont{et~al.}
  (\bibinfo{collaboration}{ATLAS}) (\bibinfo{year}{2020}{\natexlab{a}}),
  \eprint{2001.05178}.

\bibitem[{\citenamefont{Del~Duca et~al.}(2001)\citenamefont{Del~Duca, Kilgore,
  Oleari, Schmidt, and Zeppenfeld}}]{DelDuca:2001eu}
\bibinfo{author}{\bibfnamefont{V.}~\bibnamefont{Del~Duca}},
  \bibinfo{author}{\bibfnamefont{W.}~\bibnamefont{Kilgore}},
  \bibinfo{author}{\bibfnamefont{C.}~\bibnamefont{Oleari}},
  \bibinfo{author}{\bibfnamefont{C.}~\bibnamefont{Schmidt}}, \bibnamefont{and}
  \bibinfo{author}{\bibfnamefont{D.}~\bibnamefont{Zeppenfeld}},
  \bibinfo{journal}{Phys. Rev. Lett.} \textbf{\bibinfo{volume}{87}},
  \bibinfo{pages}{122001} (\bibinfo{year}{2001}), \eprint{hep-ph/0105129}.

\bibitem[{\citenamefont{Del~Duca et~al.}(2003)\citenamefont{Del~Duca, Kilgore,
  Oleari, Schmidt, and Zeppenfeld}}]{DelDuca:2003ba}
\bibinfo{author}{\bibfnamefont{V.}~\bibnamefont{Del~Duca}},
  \bibinfo{author}{\bibfnamefont{W.}~\bibnamefont{Kilgore}},
  \bibinfo{author}{\bibfnamefont{C.}~\bibnamefont{Oleari}},
  \bibinfo{author}{\bibfnamefont{C.~R.} \bibnamefont{Schmidt}},
  \bibnamefont{and}
  \bibinfo{author}{\bibfnamefont{D.}~\bibnamefont{Zeppenfeld}},
  \bibinfo{journal}{Phys. Rev.} \textbf{\bibinfo{volume}{D67}},
  \bibinfo{pages}{073003} (\bibinfo{year}{2003}), \eprint{hep-ph/0301013}.

\bibitem[{\citenamefont{del Duca et~al.}(2007)\citenamefont{del Duca, Klämke,
  Moretti, Piccinini, Pittau, Polosa, and Zeppenfeld}}]{delDuca:2007nwt}
\bibinfo{author}{\bibfnamefont{V.}~\bibnamefont{del Duca}},
  \bibinfo{author}{\bibfnamefont{G.}~\bibnamefont{Klämke}},
  \bibinfo{author}{\bibfnamefont{M.}~\bibnamefont{Moretti}},
  \bibinfo{author}{\bibfnamefont{F.}~\bibnamefont{Piccinini}},
  \bibinfo{author}{\bibfnamefont{R.}~\bibnamefont{Pittau}},
  \bibinfo{author}{\bibfnamefont{A.~D.} \bibnamefont{Polosa}},
  \bibnamefont{and}
  \bibinfo{author}{\bibfnamefont{D.}~\bibnamefont{Zeppenfeld}}, in
  \emph{\bibinfo{booktitle}{{Proceedings, 42nd Rencontres de Moriond on QCD and
  High Energy Hadronic Interactions: La Thuile, Italy, March 17-24, 2007}}},
  \bibinfo{organization}{Gioi Publ.} (\bibinfo{publisher}{Gioi Publ.},
  \bibinfo{address}{Hanoi, Vietnam}, \bibinfo{year}{2007}), pp.
  \bibinfo{pages}{205--208}.

\bibitem[{\citenamefont{Del~Duca et~al.}(2006)\citenamefont{Del~Duca, Klamke,
  Zeppenfeld, Mangano, Moretti, Piccinini, Pittau, and
  Polosa}}]{DelDuca:2006hk}
\bibinfo{author}{\bibfnamefont{V.}~\bibnamefont{Del~Duca}},
  \bibinfo{author}{\bibfnamefont{G.}~\bibnamefont{Klamke}},
  \bibinfo{author}{\bibfnamefont{D.}~\bibnamefont{Zeppenfeld}},
  \bibinfo{author}{\bibfnamefont{M.~L.} \bibnamefont{Mangano}},
  \bibinfo{author}{\bibfnamefont{M.}~\bibnamefont{Moretti}},
  \bibinfo{author}{\bibfnamefont{F.}~\bibnamefont{Piccinini}},
  \bibinfo{author}{\bibfnamefont{R.}~\bibnamefont{Pittau}}, \bibnamefont{and}
  \bibinfo{author}{\bibfnamefont{A.~D.} \bibnamefont{Polosa}},
  \bibinfo{journal}{JHEP} \textbf{\bibinfo{volume}{10}}, \bibinfo{pages}{016}
  (\bibinfo{year}{2006}), \eprint{hep-ph/0608158}.

\bibitem[{\citenamefont{Grzadkowski et~al.}(2018)\citenamefont{Grzadkowski,
  Haber, Ogreid, and Osland}}]{Grzadkowski:2018ohf}
\bibinfo{author}{\bibfnamefont{B.}~\bibnamefont{Grzadkowski}},
  \bibinfo{author}{\bibfnamefont{H.~E.} \bibnamefont{Haber}},
  \bibinfo{author}{\bibfnamefont{O.~M.} \bibnamefont{Ogreid}},
  \bibnamefont{and} \bibinfo{author}{\bibfnamefont{P.}~\bibnamefont{Osland}},
  \bibinfo{journal}{JHEP} \textbf{\bibinfo{volume}{12}}, \bibinfo{pages}{056}
  (\bibinfo{year}{2018}), \eprint{1808.01472}.

\bibitem[{\citenamefont{Fontes et~al.}(2018)\citenamefont{Fontes, Mühlleitner,
  Romão, Santos, Silva, and Wittbrodt}}]{Fontes:2017zfn}
\bibinfo{author}{\bibfnamefont{D.}~\bibnamefont{Fontes}},
  \bibinfo{author}{\bibfnamefont{M.}~\bibnamefont{Mühlleitner}},
  \bibinfo{author}{\bibfnamefont{J.~C.} \bibnamefont{Romão}},
  \bibinfo{author}{\bibfnamefont{R.}~\bibnamefont{Santos}},
  \bibinfo{author}{\bibfnamefont{J.~P.} \bibnamefont{Silva}}, \bibnamefont{and}
  \bibinfo{author}{\bibfnamefont{J.}~\bibnamefont{Wittbrodt}},
  \bibinfo{journal}{JHEP} \textbf{\bibinfo{volume}{02}}, \bibinfo{pages}{073}
  (\bibinfo{year}{2018}), \eprint{1711.09419}.

\bibitem[{\citenamefont{Georgi and Machacek}(1985)}]{Georgi:1985nv}
\bibinfo{author}{\bibfnamefont{H.}~\bibnamefont{Georgi}} \bibnamefont{and}
  \bibinfo{author}{\bibfnamefont{M.}~\bibnamefont{Machacek}},
  \bibinfo{journal}{Nucl. Phys.} \textbf{\bibinfo{volume}{B262}},
  \bibinfo{pages}{463} (\bibinfo{year}{1985}).

\bibitem[{\citenamefont{Gunion et~al.}(1990)\citenamefont{Gunion, Vega, and
  Wudka}}]{Gunion:1989ci}
\bibinfo{author}{\bibfnamefont{J.~F.} \bibnamefont{Gunion}},
  \bibinfo{author}{\bibfnamefont{R.}~\bibnamefont{Vega}}, \bibnamefont{and}
  \bibinfo{author}{\bibfnamefont{J.}~\bibnamefont{Wudka}},
  \bibinfo{journal}{Phys. Rev.} \textbf{\bibinfo{volume}{D42}},
  \bibinfo{pages}{1673} (\bibinfo{year}{1990}).

\bibitem[{\citenamefont{Hartling et~al.}(2015)\citenamefont{Hartling, Kumar,
  and Logan}}]{Hartling:2014aga}
\bibinfo{author}{\bibfnamefont{K.}~\bibnamefont{Hartling}},
  \bibinfo{author}{\bibfnamefont{K.}~\bibnamefont{Kumar}}, \bibnamefont{and}
  \bibinfo{author}{\bibfnamefont{H.~E.} \bibnamefont{Logan}},
  \bibinfo{journal}{Phys. Rev.} \textbf{\bibinfo{volume}{D91}},
  \bibinfo{pages}{015013} (\bibinfo{year}{2015}), \eprint{1410.5538}.

\bibitem[{\citenamefont{Gunion et~al.}(1991)\citenamefont{Gunion, Vega, and
  Wudka}}]{Gunion:1990dt}
\bibinfo{author}{\bibfnamefont{J.~F.} \bibnamefont{Gunion}},
  \bibinfo{author}{\bibfnamefont{R.}~\bibnamefont{Vega}}, \bibnamefont{and}
  \bibinfo{author}{\bibfnamefont{J.}~\bibnamefont{Wudka}},
  \bibinfo{journal}{Phys. Rev.} \textbf{\bibinfo{volume}{D43}},
  \bibinfo{pages}{2322} (\bibinfo{year}{1991}).

\bibitem[{\citenamefont{Ferretti}(2014)}]{Ferretti:2014qta}
\bibinfo{author}{\bibfnamefont{G.}~\bibnamefont{Ferretti}},
  \bibinfo{journal}{JHEP} \textbf{\bibinfo{volume}{06}}, \bibinfo{pages}{142}
  (\bibinfo{year}{2014}), \eprint{1404.7137}.

\bibitem[{\citenamefont{Golterman and Shamir}(2015)}]{Golterman:2015zwa}
\bibinfo{author}{\bibfnamefont{M.}~\bibnamefont{Golterman}} \bibnamefont{and}
  \bibinfo{author}{\bibfnamefont{Y.}~\bibnamefont{Shamir}},
  \bibinfo{journal}{Phys. Rev.} \textbf{\bibinfo{volume}{D91}},
  \bibinfo{pages}{094506} (\bibinfo{year}{2015}), \eprint{1502.00390}.

\bibitem[{\citenamefont{Godfrey and Moats}(2010)}]{Godfrey:2010qb}
\bibinfo{author}{\bibfnamefont{S.}~\bibnamefont{Godfrey}} \bibnamefont{and}
  \bibinfo{author}{\bibfnamefont{K.}~\bibnamefont{Moats}},
  \bibinfo{journal}{Phys. Rev.} \textbf{\bibinfo{volume}{D81}},
  \bibinfo{pages}{075026} (\bibinfo{year}{2010}), \eprint{1003.3033}.

\bibitem[{\citenamefont{Cheung and Ghosh}(2002)}]{Cheung:2002gd}
\bibinfo{author}{\bibfnamefont{K.}~\bibnamefont{Cheung}} \bibnamefont{and}
  \bibinfo{author}{\bibfnamefont{D.~K.} \bibnamefont{Ghosh}},
  \bibinfo{journal}{JHEP} \textbf{\bibinfo{volume}{11}}, \bibinfo{pages}{048}
  (\bibinfo{year}{2002}), \eprint{hep-ph/0208254}.

\bibitem[{\citenamefont{Englert et~al.}(2013)\citenamefont{Englert, Re, and
  Spannowsky}}]{Englert:2013wga}
\bibinfo{author}{\bibfnamefont{C.}~\bibnamefont{Englert}},
  \bibinfo{author}{\bibfnamefont{E.}~\bibnamefont{Re}}, \bibnamefont{and}
  \bibinfo{author}{\bibfnamefont{M.}~\bibnamefont{Spannowsky}},
  \bibinfo{journal}{Phys. Rev.} \textbf{\bibinfo{volume}{D88}},
  \bibinfo{pages}{035024} (\bibinfo{year}{2013}), \eprint{1306.6228}.

\bibitem[{\citenamefont{Zaro and Logan}(2015)}]{Zaro:2015ika}
\bibinfo{author}{\bibfnamefont{M.}~\bibnamefont{Zaro}} \bibnamefont{and}
  \bibinfo{author}{\bibfnamefont{H.}~\bibnamefont{Logan}}
  (\bibinfo{year}{2015}).

\bibitem[{\citenamefont{Degrande et~al.}(2016)\citenamefont{Degrande, Hartling,
  Logan, Peterson, and Zaro}}]{Degrande:2015xnm}
\bibinfo{author}{\bibfnamefont{C.}~\bibnamefont{Degrande}},
  \bibinfo{author}{\bibfnamefont{K.}~\bibnamefont{Hartling}},
  \bibinfo{author}{\bibfnamefont{H.~E.} \bibnamefont{Logan}},
  \bibinfo{author}{\bibfnamefont{A.~D.} \bibnamefont{Peterson}},
  \bibnamefont{and} \bibinfo{author}{\bibfnamefont{M.}~\bibnamefont{Zaro}},
  \bibinfo{journal}{Phys. Rev.} \textbf{\bibinfo{volume}{D93}},
  \bibinfo{pages}{035004} (\bibinfo{year}{2016}), \eprint{1512.01243}.

\bibitem[{\citenamefont{Binoth and van~der Bij}(1997)}]{Binoth:1996au}
\bibinfo{author}{\bibfnamefont{T.}~\bibnamefont{Binoth}} \bibnamefont{and}
  \bibinfo{author}{\bibfnamefont{J.~J.} \bibnamefont{van~der Bij}},
  \bibinfo{journal}{Z. Phys.} \textbf{\bibinfo{volume}{C75}},
  \bibinfo{pages}{17} (\bibinfo{year}{1997}), \eprint{hep-ph/9608245}.

\bibitem[{\citenamefont{Dittmaier et~al.}(2011)}]{Dittmaier:2011ti}
\bibinfo{author}{\bibfnamefont{S.}~\bibnamefont{Dittmaier}}
  \bibnamefont{et~al.} (\bibinfo{collaboration}{LHC Higgs Cross Section Working
  Group}) (\bibinfo{year}{2011}), \eprint{1101.0593}.

\bibitem[{\citenamefont{Das}(2019)}]{Das:2018fog}
\bibinfo{author}{\bibfnamefont{D.}~\bibnamefont{Das}}, \bibinfo{journal}{Phys.
  Rev. D} \textbf{\bibinfo{volume}{99}}, \bibinfo{pages}{095035}
  (\bibinfo{year}{2019}), \eprint{1804.06630}.

\bibitem[{\citenamefont{Gaemers and Hoogeveen}(1984)}]{Gaemers:1984sj}
\bibinfo{author}{\bibfnamefont{K.~J.~F.} \bibnamefont{Gaemers}}
  \bibnamefont{and}
  \bibinfo{author}{\bibfnamefont{F.}~\bibnamefont{Hoogeveen}},
  \bibinfo{journal}{Phys. Lett.} \textbf{\bibinfo{volume}{146B}},
  \bibinfo{pages}{347} (\bibinfo{year}{1984}).

\bibitem[{\citenamefont{Dicus et~al.}(1994)\citenamefont{Dicus, Stange, and
  Willenbrock}}]{Dicus:1994bm}
\bibinfo{author}{\bibfnamefont{D.}~\bibnamefont{Dicus}},
  \bibinfo{author}{\bibfnamefont{A.}~\bibnamefont{Stange}}, \bibnamefont{and}
  \bibinfo{author}{\bibfnamefont{S.}~\bibnamefont{Willenbrock}},
  \bibinfo{journal}{Phys. Lett.} \textbf{\bibinfo{volume}{B333}},
  \bibinfo{pages}{126} (\bibinfo{year}{1994}), \eprint{hep-ph/9404359}.

\bibitem[{\citenamefont{Frederix and Maltoni}(2009)}]{Frederix:2007gi}
\bibinfo{author}{\bibfnamefont{R.}~\bibnamefont{Frederix}} \bibnamefont{and}
  \bibinfo{author}{\bibfnamefont{F.}~\bibnamefont{Maltoni}},
  \bibinfo{journal}{JHEP} \textbf{\bibinfo{volume}{01}}, \bibinfo{pages}{047}
  (\bibinfo{year}{2009}), \eprint{0712.2355}.

\bibitem[{\citenamefont{Carena and Liu}(2016)}]{Carena:2016npr}
\bibinfo{author}{\bibfnamefont{M.}~\bibnamefont{Carena}} \bibnamefont{and}
  \bibinfo{author}{\bibfnamefont{Z.}~\bibnamefont{Liu}},
  \bibinfo{journal}{JHEP} \textbf{\bibinfo{volume}{11}}, \bibinfo{pages}{159}
  (\bibinfo{year}{2016}), \eprint{1608.07282}.

\bibitem[{\citenamefont{Hespel et~al.}(2016)\citenamefont{Hespel, Maltoni, and
  Vryonidou}}]{Hespel:2016qaf}
\bibinfo{author}{\bibfnamefont{B.}~\bibnamefont{Hespel}},
  \bibinfo{author}{\bibfnamefont{F.}~\bibnamefont{Maltoni}}, \bibnamefont{and}
  \bibinfo{author}{\bibfnamefont{E.}~\bibnamefont{Vryonidou}},
  \bibinfo{journal}{JHEP} \textbf{\bibinfo{volume}{10}}, \bibinfo{pages}{016}
  (\bibinfo{year}{2016}), \eprint{1606.04149}.

\bibitem[{\citenamefont{Buarque~Franzosi
  et~al.}(2018)\citenamefont{Buarque~Franzosi, Fabbri, and
  Schumann}}]{BuarqueFranzosi:2017qlm}
\bibinfo{author}{\bibfnamefont{D.}~\bibnamefont{Buarque~Franzosi}},
  \bibinfo{author}{\bibfnamefont{F.}~\bibnamefont{Fabbri}}, \bibnamefont{and}
  \bibinfo{author}{\bibfnamefont{S.}~\bibnamefont{Schumann}},
  \bibinfo{journal}{JHEP} \textbf{\bibinfo{volume}{03}}, \bibinfo{pages}{022}
  (\bibinfo{year}{2018}), \eprint{1711.00102}.

\bibitem[{\citenamefont{Englert et~al.}(2020)\citenamefont{Englert, Galler, and
  White}}]{Englert:2019rga}
\bibinfo{author}{\bibfnamefont{C.}~\bibnamefont{Englert}},
  \bibinfo{author}{\bibfnamefont{P.}~\bibnamefont{Galler}}, \bibnamefont{and}
  \bibinfo{author}{\bibfnamefont{C.~D.} \bibnamefont{White}},
  \bibinfo{journal}{Phys. Rev.} \textbf{\bibinfo{volume}{D101}},
  \bibinfo{pages}{035035} (\bibinfo{year}{2020}), \eprint{1908.05588}.

\bibitem[{\citenamefont{Basler et~al.}(2020)\citenamefont{Basler, Dawson,
  Englert, and Mühlleitner}}]{Basler:2019nas}
\bibinfo{author}{\bibfnamefont{P.}~\bibnamefont{Basler}},
  \bibinfo{author}{\bibfnamefont{S.}~\bibnamefont{Dawson}},
  \bibinfo{author}{\bibfnamefont{C.}~\bibnamefont{Englert}}, \bibnamefont{and}
  \bibinfo{author}{\bibfnamefont{M.}~\bibnamefont{Mühlleitner}},
  \bibinfo{journal}{Phys. Rev.} \textbf{\bibinfo{volume}{D101}},
  \bibinfo{pages}{015019} (\bibinfo{year}{2020}), \eprint{1909.09987}.

\bibitem[{\citenamefont{Chen et~al.}(2015)\citenamefont{Chen, Dawson, and
  Lewis}}]{Chen:2014ask}
\bibinfo{author}{\bibfnamefont{C.-Y.} \bibnamefont{Chen}},
  \bibinfo{author}{\bibfnamefont{S.}~\bibnamefont{Dawson}}, \bibnamefont{and}
  \bibinfo{author}{\bibfnamefont{I.~M.} \bibnamefont{Lewis}},
  \bibinfo{journal}{Phys. Rev.} \textbf{\bibinfo{volume}{D91}},
  \bibinfo{pages}{035015} (\bibinfo{year}{2015}), \eprint{1410.5488}.

\bibitem[{\citenamefont{Bojarski et~al.}(2016)\citenamefont{Bojarski, Chalons,
  Lopez-Val, and Robens}}]{Bojarski:2015kra}
\bibinfo{author}{\bibfnamefont{F.}~\bibnamefont{Bojarski}},
  \bibinfo{author}{\bibfnamefont{G.}~\bibnamefont{Chalons}},
  \bibinfo{author}{\bibfnamefont{D.}~\bibnamefont{Lopez-Val}},
  \bibnamefont{and} \bibinfo{author}{\bibfnamefont{T.}~\bibnamefont{Robens}},
  \bibinfo{journal}{JHEP} \textbf{\bibinfo{volume}{02}}, \bibinfo{pages}{147}
  (\bibinfo{year}{2016}), \eprint{1511.08120}.

\bibitem[{\citenamefont{Dawson and Sullivan}(2018)}]{Dawson:2017jja}
\bibinfo{author}{\bibfnamefont{S.}~\bibnamefont{Dawson}} \bibnamefont{and}
  \bibinfo{author}{\bibfnamefont{M.}~\bibnamefont{Sullivan}},
  \bibinfo{journal}{Phys. Rev.} \textbf{\bibinfo{volume}{D97}},
  \bibinfo{pages}{015022} (\bibinfo{year}{2018}), \eprint{1711.06683}.

\bibitem[{\citenamefont{López-Val and Robens}(2014)}]{Lopez-Val:2014jva}
\bibinfo{author}{\bibfnamefont{D.}~\bibnamefont{López-Val}} \bibnamefont{and}
  \bibinfo{author}{\bibfnamefont{T.}~\bibnamefont{Robens}},
  \bibinfo{journal}{Phys. Rev.} \textbf{\bibinfo{volume}{D90}},
  \bibinfo{pages}{114018} (\bibinfo{year}{2014}), \eprint{1406.1043}.

\bibitem[{\citenamefont{Falkowski et~al.}(2015)\citenamefont{Falkowski, Gross,
  and Lebedev}}]{Falkowski:2015iwa}
\bibinfo{author}{\bibfnamefont{A.}~\bibnamefont{Falkowski}},
  \bibinfo{author}{\bibfnamefont{C.}~\bibnamefont{Gross}}, \bibnamefont{and}
  \bibinfo{author}{\bibfnamefont{O.}~\bibnamefont{Lebedev}},
  \bibinfo{journal}{JHEP} \textbf{\bibinfo{volume}{05}}, \bibinfo{pages}{057}
  (\bibinfo{year}{2015}), \eprint{1502.01361}.

\bibitem[{\citenamefont{Robens and Stefaniak}(2015)}]{Robens:2015gla}
\bibinfo{author}{\bibfnamefont{T.}~\bibnamefont{Robens}} \bibnamefont{and}
  \bibinfo{author}{\bibfnamefont{T.}~\bibnamefont{Stefaniak}},
  \bibinfo{journal}{Eur. Phys. J.} \textbf{\bibinfo{volume}{C75}},
  \bibinfo{pages}{104} (\bibinfo{year}{2015}), \eprint{1501.02234}.

\bibitem[{\citenamefont{Arnold et~al.}(2009)}]{Arnold:2008rz}
\bibinfo{author}{\bibfnamefont{K.}~\bibnamefont{Arnold}} \bibnamefont{et~al.},
  \bibinfo{journal}{Comput. Phys. Commun.} \textbf{\bibinfo{volume}{180}},
  \bibinfo{pages}{1661} (\bibinfo{year}{2009}), \eprint{0811.4559}.

\bibitem[{\citenamefont{Alwall et~al.}(2014)\citenamefont{Alwall, Frederix,
  Frixione, Hirschi, Maltoni, Mattelaer, Shao, Stelzer, Torrielli, and
  Zaro}}]{Alwall:2014hca}
\bibinfo{author}{\bibfnamefont{J.}~\bibnamefont{Alwall}},
  \bibinfo{author}{\bibfnamefont{R.}~\bibnamefont{Frederix}},
  \bibinfo{author}{\bibfnamefont{S.}~\bibnamefont{Frixione}},
  \bibinfo{author}{\bibfnamefont{V.}~\bibnamefont{Hirschi}},
  \bibinfo{author}{\bibfnamefont{F.}~\bibnamefont{Maltoni}},
  \bibinfo{author}{\bibfnamefont{O.}~\bibnamefont{Mattelaer}},
  \bibinfo{author}{\bibfnamefont{H.~S.} \bibnamefont{Shao}},
  \bibinfo{author}{\bibfnamefont{T.}~\bibnamefont{Stelzer}},
  \bibinfo{author}{\bibfnamefont{P.}~\bibnamefont{Torrielli}},
  \bibnamefont{and} \bibinfo{author}{\bibfnamefont{M.}~\bibnamefont{Zaro}},
  \bibinfo{journal}{JHEP} \textbf{\bibinfo{volume}{07}}, \bibinfo{pages}{079}
  (\bibinfo{year}{2014}), \eprint{1405.0301}.

\bibitem[{\citenamefont{Sjostrand et~al.}(2008)\citenamefont{Sjostrand, Mrenna,
  and Skands}}]{Sjostrand:2007gs}
\bibinfo{author}{\bibfnamefont{T.}~\bibnamefont{Sjostrand}},
  \bibinfo{author}{\bibfnamefont{S.}~\bibnamefont{Mrenna}}, \bibnamefont{and}
  \bibinfo{author}{\bibfnamefont{P.~Z.} \bibnamefont{Skands}},
  \bibinfo{journal}{Comput. Phys. Commun.} \textbf{\bibinfo{volume}{178}},
  \bibinfo{pages}{852} (\bibinfo{year}{2008}), \eprint{0710.3820}.

\bibitem[{\citenamefont{Cacciari et~al.}(2012)\citenamefont{Cacciari, Salam,
  and Soyez}}]{Cacciari:2011ma}
\bibinfo{author}{\bibfnamefont{M.}~\bibnamefont{Cacciari}},
  \bibinfo{author}{\bibfnamefont{G.~P.} \bibnamefont{Salam}}, \bibnamefont{and}
  \bibinfo{author}{\bibfnamefont{G.}~\bibnamefont{Soyez}},
  \bibinfo{journal}{Eur. Phys. J.} \textbf{\bibinfo{volume}{C72}},
  \bibinfo{pages}{1896} (\bibinfo{year}{2012}), \eprint{1111.6097}.

\bibitem[{\citenamefont{Buccioni et~al.}(2019)\citenamefont{Buccioni, Kallweit,
  Pozzorini, and Zoller}}]{Buccioni:2019plc}
\bibinfo{author}{\bibfnamefont{F.}~\bibnamefont{Buccioni}},
  \bibinfo{author}{\bibfnamefont{S.}~\bibnamefont{Kallweit}},
  \bibinfo{author}{\bibfnamefont{S.}~\bibnamefont{Pozzorini}},
  \bibnamefont{and} \bibinfo{author}{\bibfnamefont{M.~F.}
  \bibnamefont{Zoller}}, \bibinfo{journal}{JHEP} \textbf{\bibinfo{volume}{12}},
  \bibinfo{pages}{015} (\bibinfo{year}{2019}), \eprint{1907.13624}.

\bibitem[{\citenamefont{Figy et~al.}(2003)\citenamefont{Figy, Oleari, and
  Zeppenfeld}}]{Figy:2003nv}
\bibinfo{author}{\bibfnamefont{T.}~\bibnamefont{Figy}},
  \bibinfo{author}{\bibfnamefont{C.}~\bibnamefont{Oleari}}, \bibnamefont{and}
  \bibinfo{author}{\bibfnamefont{D.}~\bibnamefont{Zeppenfeld}},
  \bibinfo{journal}{Phys. Rev.} \textbf{\bibinfo{volume}{D68}},
  \bibinfo{pages}{073005} (\bibinfo{year}{2003}), \eprint{hep-ph/0306109}.

\bibitem[{\citenamefont{Campbell et~al.}(2006)\citenamefont{Campbell, Ellis,
  and Zanderighi}}]{Campbell:2006xx}
\bibinfo{author}{\bibfnamefont{J.~M.} \bibnamefont{Campbell}},
  \bibinfo{author}{\bibfnamefont{R.~K.} \bibnamefont{Ellis}}, \bibnamefont{and}
  \bibinfo{author}{\bibfnamefont{G.}~\bibnamefont{Zanderighi}},
  \bibinfo{journal}{JHEP} \textbf{\bibinfo{volume}{10}}, \bibinfo{pages}{028}
  (\bibinfo{year}{2006}), \eprint{hep-ph/0608194}.

\bibitem[{\citenamefont{Campbell and Neumann}(2019)}]{Campbell:2019dru}
\bibinfo{author}{\bibfnamefont{J.}~\bibnamefont{Campbell}} \bibnamefont{and}
  \bibinfo{author}{\bibfnamefont{T.}~\bibnamefont{Neumann}},
  \bibinfo{journal}{JHEP} \textbf{\bibinfo{volume}{12}}, \bibinfo{pages}{034}
  (\bibinfo{year}{2019}), \eprint{1909.09117}.

\bibitem[{\citenamefont{Derrick}(1987)}]{Derrick:1987uy}
\bibinfo{editor}{\bibfnamefont{M.}~\bibnamefont{Derrick}}, ed.,
  \emph{\bibinfo{title}{{PHYSICS IN COLLISION 6. PROCEEDINGS, 6TH INTERNATIONAL
  CONFERENCE, CHICAGO, USA, SEPTEMBER 3-5, 1986}}} (\bibinfo{year}{1987}).

\bibitem[{\citenamefont{Bjorken}(1993)}]{Bjorken:1992er}
\bibinfo{author}{\bibfnamefont{J.~D.} \bibnamefont{Bjorken}},
  \bibinfo{journal}{Phys. Rev.} \textbf{\bibinfo{volume}{D47}},
  \bibinfo{pages}{101} (\bibinfo{year}{1993}).

\bibitem[{\citenamefont{Barger et~al.}(1991)\citenamefont{Barger, Cheung, Han,
  and Zeppenfeld}}]{Barger:1991ar}
\bibinfo{author}{\bibfnamefont{V.~D.} \bibnamefont{Barger}},
  \bibinfo{author}{\bibfnamefont{K.-m.} \bibnamefont{Cheung}},
  \bibinfo{author}{\bibfnamefont{T.}~\bibnamefont{Han}}, \bibnamefont{and}
  \bibinfo{author}{\bibfnamefont{D.}~\bibnamefont{Zeppenfeld}},
  \bibinfo{journal}{Phys. Rev.} \textbf{\bibinfo{volume}{D44}},
  \bibinfo{pages}{2701} (\bibinfo{year}{1991}), \bibinfo{note}{[Erratum: Phys.
  Rev.D48,5444(1993)]}.

\bibitem[{\citenamefont{Barger et~al.}(1995)\citenamefont{Barger, Phillips, and
  Zeppenfeld}}]{Barger:1994zq}
\bibinfo{author}{\bibfnamefont{V.~D.} \bibnamefont{Barger}},
  \bibinfo{author}{\bibfnamefont{R.~J.~N.} \bibnamefont{Phillips}},
  \bibnamefont{and}
  \bibinfo{author}{\bibfnamefont{D.}~\bibnamefont{Zeppenfeld}},
  \bibinfo{journal}{Phys. Lett.} \textbf{\bibinfo{volume}{B346}},
  \bibinfo{pages}{106} (\bibinfo{year}{1995}), \eprint{hep-ph/9412276}.

\bibitem[{\citenamefont{Dawson}(1985)}]{Dawson:1984gx}
\bibinfo{author}{\bibfnamefont{S.}~\bibnamefont{Dawson}},
  \bibinfo{journal}{Nucl. Phys.} \textbf{\bibinfo{volume}{B249}},
  \bibinfo{pages}{42} (\bibinfo{year}{1985}).

\bibitem[{\citenamefont{Figy et~al.}(2008)\citenamefont{Figy, Hankele, and
  Zeppenfeld}}]{Figy:2007kv}
\bibinfo{author}{\bibfnamefont{T.}~\bibnamefont{Figy}},
  \bibinfo{author}{\bibfnamefont{V.}~\bibnamefont{Hankele}}, \bibnamefont{and}
  \bibinfo{author}{\bibfnamefont{D.}~\bibnamefont{Zeppenfeld}},
  \bibinfo{journal}{JHEP} \textbf{\bibinfo{volume}{02}}, \bibinfo{pages}{076}
  (\bibinfo{year}{2008}), \eprint{0710.5621}.

\bibitem[{\citenamefont{Goncalves et~al.}(2017)\citenamefont{Goncalves, Plehn,
  and Thompson}}]{Goncalves:2017gzy}
\bibinfo{author}{\bibfnamefont{D.}~\bibnamefont{Goncalves}},
  \bibinfo{author}{\bibfnamefont{T.}~\bibnamefont{Plehn}}, \bibnamefont{and}
  \bibinfo{author}{\bibfnamefont{J.~M.} \bibnamefont{Thompson}},
  \bibinfo{journal}{Phys. Rev.} \textbf{\bibinfo{volume}{D95}},
  \bibinfo{pages}{095011} (\bibinfo{year}{2017}), \eprint{1702.05098}.

\bibitem[{\citenamefont{Sirunyan et~al.}(2018)}]{Sirunyan:2018zkk}
\bibinfo{author}{\bibfnamefont{A.~M.} \bibnamefont{Sirunyan}}
  \bibnamefont{et~al.} (\bibinfo{collaboration}{CMS}), \bibinfo{journal}{JHEP}
  \textbf{\bibinfo{volume}{08}}, \bibinfo{pages}{152} (\bibinfo{year}{2018}),
  \eprint{1806.03548}.

\bibitem[{\citenamefont{Aaboud et~al.}(2019)}]{Aaboud:2018knk}
\bibinfo{author}{\bibfnamefont{M.}~\bibnamefont{Aaboud}} \bibnamefont{et~al.}
  (\bibinfo{collaboration}{ATLAS}), \bibinfo{journal}{JHEP}
  \textbf{\bibinfo{volume}{01}}, \bibinfo{pages}{030} (\bibinfo{year}{2019}),
  \eprint{1804.06174}.

\bibitem[{\citenamefont{Aaboud et~al.}(2018)}]{Aaboud:2018urx}
\bibinfo{author}{\bibfnamefont{M.}~\bibnamefont{Aaboud}} \bibnamefont{et~al.}
  (\bibinfo{collaboration}{ATLAS}), \bibinfo{journal}{Phys. Lett.}
  \textbf{\bibinfo{volume}{B784}}, \bibinfo{pages}{173} (\bibinfo{year}{2018}),
  \eprint{1806.00425}.

\bibitem[{\citenamefont{Aad et~al.}(2020{\natexlab{b}})}]{Aad:2020kop}
\bibinfo{author}{\bibfnamefont{G.}~\bibnamefont{Aad}} \bibnamefont{et~al.}
  (\bibinfo{collaboration}{ATLAS}) (\bibinfo{year}{2020}{\natexlab{b}}),
  \eprint{2005.05138}.

\bibitem[{\citenamefont{Read}(2002)}]{Read:2002hq}
\bibinfo{author}{\bibfnamefont{A.~L.} \bibnamefont{Read}}, \bibinfo{journal}{J.
  Phys.} \textbf{\bibinfo{volume}{G28}}, \bibinfo{pages}{2693}
  (\bibinfo{year}{2002}).

\bibitem[{\citenamefont{Andersen et~al.}(2013)}]{Heinemeyer:2013tqa}
\bibinfo{author}{\bibfnamefont{J.~R.} \bibnamefont{Andersen}}
  \bibnamefont{et~al.} (\bibinfo{collaboration}{LHC Higgs Cross Section Working
  Group}) (\bibinfo{year}{2013}), \eprint{1307.1347}.

\bibitem[{\citenamefont{de~Blas et~al.}(2020)}]{deBlas:2019rxi}
\bibinfo{author}{\bibfnamefont{J.}~\bibnamefont{de~Blas}} \bibnamefont{et~al.},
  \bibinfo{journal}{JHEP} \textbf{\bibinfo{volume}{01}}, \bibinfo{pages}{139}
  (\bibinfo{year}{2020}), \eprint{1905.03764}.

\end{thebibliography}

\end{document}